%Paper: hep-th/9412150
%From: Ramzi Khuri <rrk@hep.physics.mcgill.ca>
%Date: Fri, 16 Dec 1994 17:06:14 -0500 (EST)

\headline={\ifnum\pageno=1\firstheadline\else
\ifodd\pageno\rightheadline \else\leftheadline\fi\fi}
\def\firstheadline{\hfil}
\def\rightheadline{\hfil}
\def\leftheadline{\hfil}

\footline={\ifnum\pageno>0 \hss \folio \hss \else\fi}

%\footline={\ifnum\pageno=1\firstfootline\else\otherfootline\fi}

\font\tenrm=cmr10

\font\elevenbf=cmbx10 scaled\magstep 1
\font\elevenrm=cmr10 scaled\magstep 1
\font\elevenit=cmti10 scaled\magstep 1

%\nopagenumbers
\hsize=6.0truein
\vsize=8.5truein
\parindent=1.5pc
\baselineskip=10pt
%% FOLLOWING LINE CANNOT BE BROKEN BEFORE 70 CHAR
%==========================================================================
\rightline{CERN-TH.7467/94}
\rightline{McGill/94--51}
\centerline{\elevenbf Black Hole Solutions in String
Theory}\footnote{}
{Based on talks given at MRST 94, McGill University, May 1994
and
at Mini-workshop on Extended Objects, Newton Institute,
March 1994
and condensed from work already published$^{1,2}$.}
%% FOLLOWING LINE CANNOT BE BROKEN BEFORE 70 CHAR
%==========================================================================
\vglue 1.0cm
\centerline{\elevenrm Ramzi R. Khuri\footnote{$^\dagger$}
{Supported by a World Laboratory Fellowship.}}
\baselineskip=13pt
\centerline{\elevenit CERN, Theory Division}
\baselineskip=12pt
\centerline{\elevenit CH-1211 Geneva 23, Switzerland}
\baselineskip=14pt
\centerline{\elevenrm and}
\baselineskip=14pt
\centerline{\elevenit McGill University, Physics Department}
\baselineskip=12pt
\centerline{\elevenit Montreal, PQ, H3A 2T8, Canada}
%% FOLLOWING LINE CANNOT BE BROKEN BEFORE 70 CHAR
%==========================================================================
\vglue 0.8cm
\centerline{\tenrm ABSTRACT}
\vglue 0.3cm
  {\rightskip=3pc
 \leftskip=3pc
 \tenrm\baselineskip=12pt
 \noindent
We present two-parameter solutions of the low-energy
four-dimensional
heterotic string which in the extremal limit
reduce to supersymmetric monopole,
string and domain wall solutions. The effective scalar
coupling to the Maxwell field, $e^{-\alpha \phi} F_{\mu\nu}
F^{\mu\nu}$, gives
rise to a new string black hole with $\alpha = \sqrt{3}$, in
contrast
to the
pure dilaton black hole solution which has $\alpha=1$.
Implications of
string/fivebrane duality in $D=10$ to
four-dimensional dualities are discussed.
\vglue 0.8cm }
%% FOLLOWING LINE CANNOT BE BROKEN BEFORE 70 CHAR
%==========================================================================
\baselineskip=14pt
\elevenrm
In recent work$^1$, supersymmetric soliton solutions of the
four-dimensional heterotic string were presented, describing
monopoles,
strings and domain walls. These solutions admit the $D=10$
interpretation of
a fivebrane wrapped around 5, 4 or 3 of the 6 compactified
dimensions and are arguably exact to all orders in $\alpha'$.
In this
talk,
we extend all three solutions to two-parameter solutions of the
low-energy
equations of the four-dimensional heterotic string$^2$. The
two-parameter solution extending the supersymmetric monopole
corresponds to
a magnetically charged black hole, while the solution extending
the
supersymmetric domain wall corresponds to a black membrane. By
contrast, the
two-parameter string solution does not possess a finite horizon
and
corresponds to a naked singularity.

All three solutions involve both the
dilaton and the modulus fields, and are thus to be contrasted
with
pure dilaton
solutions$^3$. In particular, the effective scalar coupling to
the
Maxwell field,
$e^{-\alpha\phi} F_{\mu\nu} F^{\mu\nu}$, gives rise to a new
string
black hole
with $\alpha = \sqrt{3}$, in contrast to the pure dilaton
solution of
the
heterotic string which has $\alpha = 1$ $^3$. It thus resembles
the black
hole arising from Kaluza-Klein theories
which also has $\alpha = \sqrt{3}$, and which reduces to the
Pollard-Gross-Perry-Sorkin$^4$ magnetic monopole in the extremal
limit.
The fact that the heterotic string admits $\alpha = \sqrt{3}$
black
holes also
has implications for string/fivebrane duality$^5$. Both
electric/magnetic duality and string/string duality
in $D=4$ may be seen as a consequence of string/fivebrane
duality in $D=10$.

We begin with the two-parameter black hole.  Inspired by the
wrapping
of a fivebrane around five  of the six compactified dimensions
$(x_5, x_6, x_7, x_8, x_9)$, it was shown$^1$ that the
tree-level
effective action for the $D=10$ heterotic string may be reduced
to
the
following four-dimensional form
$$ S_{1}={1\over 2\kappa^2}\int d^4 x \sqrt{-g} e^{-2\Phi -
\sigma_{1}}
\left( R + 4(\partial\Phi)^2 + 4\partial\sigma_{1}\cdot
\partial\Phi -
{1\over 4} e^{2\sigma_{1}} F_{\mu\nu} F^{\mu\nu}\right),
\eqno(1) $$
where $\mu,\nu=0,1,2,3$. Here $g_{\mu\nu}$ is the string
sigma-model metric and $\Phi$ is the dilaton. In the case of
toroidal
compactification, with $N=4$ supersymmetry in $D=4$,
$\sigma_{1}$ is
a modulus
field, $g_{44}=e^{-2 \sigma_1}$, and
$F_{\mu\nu}=H_{\mu\nu4}$ where $H=dB$ and $B$ is the
string antisymmetric tensor. However, actions of this type also
appear in a
large class of $N=1$ supergravity theories$^6$. The solution is
given by$^2$
$$ \eqalignno{e^{-2\Phi}&=e^{2\sigma_{1}}=\left(1 - {r_-\over r}
\right),\cr
ds^2&=-\left(1-{r_+\over r}\right)\left(1-{r_-\over
r}\right)^{-1}dt^2 +
\left(1-{r_+\over r}\right)^{-1}dr^2 +
r^2\left(1-{r_-\over r}\right)d\Omega_2^2,\cr
F_{\theta\varphi}&=\sqrt{r_+r_-} \sin\theta, &(2)\cr} $$
where here, and throughout this paper, we set the dilaton vev
$\Phi_0$ equal to
zero. This represents a magnetically charged black hole with
event horizon at $r=r_+$ and inner horizon at $r=r_-$.
The magnetic charge and mass of the black hole are given by
$$ \eqalignno{g_{1}&={4\pi\over
{\sqrt{2}\kappa}}(r_+r_-)^{{1}\over{2}},
   \cr
{\cal M}_1&={{2 \pi}\over {{\kappa}^2}}(2r_+ - r_-). &(3)\cr} $$
Changing coordinates via $y=r-r_-$ and taking the extremal limit
$r_+=r_-$
yields:
$$ \eqalignno{e^{2\Phi}&=e^{-2\sigma_{1}}=\left(1 + {r_-\over y}
\right),\cr
ds^2&=-dt^2 + e^{2\Phi}\left(dy^2 + y^2d\Omega_2^2\right),\cr
F_{\theta\varphi}&=r_- \sin\theta, &(4)\cr} $$
which is just the tree-level supersymmetric monopole solution
without
a Yang-Mills field which saturates the Bogomol'nyi bound
$\sqrt{2}
\kappa{\cal M}_1\ge g_1$. Note that the monopole arises in the
gravitational sector of the string, as can be seen from an
earlier
solution found in purely bosonic string theory$^7$.
Supersymmetric
extensions of this solution both with and without gauge fields
were
found in $^8$.

Next we derive a two-parameter string solution which, however,
does not possess a finite event horizon and consequently cannot
be
interpreted as a black string. This is inspired by the wrapping
 of
the
fivebrane around four of the compactified dimensions $(x_6, x_7,
x_8,x_9)$.
The action is given by
$$ S_{2}={1\over 2\kappa^2}\int d^4 x \sqrt{-g} e^{-2\Phi -
2\sigma_{2}}
\left( R + 4(\partial\Phi)^2 + 8\partial\sigma_{2}
\cdot\partial\Phi +
2(\partial\sigma_{2})^2 - {1\over 2} e^{4\sigma_{2}}
F_\mu F^\mu
\right), \eqno(5) $$
In the case of the torus, $\sigma_2$ is the modulus field
$g_{44}=g_{55}=e^{-2\sigma_2}$ and
$F_{\mu}=H_{\mu45}$. A two-parameter family of solutions is now
given by$^2$
$$ \eqalignno{e^{2\Phi}&=e^{-2\sigma_{2}}=
(1+k/2-\lambda \ln y),\cr
ds^2&=-(1+k)dt^2 + (1+k)^{-1}(1+k/2-\lambda \ln y)dy^2 +
y^2(1+k/2-\lambda \ln y)d\theta^2 + dx_3^2,\cr
F_\theta&=\lambda \sqrt{1+k}, &(6)\cr} $$
where for $k=0$ we recover the supersymmetric string soliton
solution$^1$ which is dual to the elementary string solution of
Dabholkar {\it et al}$^9$. The solution shown in Eq.(6) in fact
represents a naked singularity,
since the event horizon is pushed out to $r_+=\infty$, which
agrees
with
the Horowitz-Strominger ``no-$4D$-black-string'' theorem$^{10}$.

Finally, we consider the two-parameter black membrane solution.
In
this case,
we wrap the fivebrane around three of the compactified
dimensions
$(x_7, x_8,
x_9)$.
However, the four-dimensional action necessary to yield membrane
solutions
is not obtained by a simple dimensional reduction of the
ten-dimensional action because of the non-vanishing of
$F=H_{456}$.
Instead, the effective action is obtained by treating $F^2$ as a
cosmological
constant and is  given by
$$ S_{3}={1\over 2\kappa^2}\int d^4 x \sqrt{-g}
e^{-2\Phi - 3\sigma_{3}}
\left( R + 4(\partial\Phi)^2 + 12\partial\sigma_{3}
\cdot\partial\Phi
+
6(\partial\sigma_{3})^2 - e^{6\sigma_{3}} {1\over 2} F^2 \right),
\eqno(7) $$
In the case of the torus, $\sigma_3$ is the modulus field
$g_{44}=g_{55}=g_{66}=e^{-2\sigma_3}$.
The two-parameter black membrane solution is then$^2$
$$ \eqalignno{e^{-2\Phi}&=e^{2\sigma_{3}}=\left(1-{r\over
r_-}\right),\cr
ds^2&=-\left(1-{r\over r_+}\right)\left(1-{r\over
r_-}\right)^{-1}dt^2 +
\left(1-{r\over r_+}\right)^{-1}\left(1-{r\over r_-}
\right)^{-4}dr^2
+
dx_2^2 + dx_3^2,\cr F&=-(r_+r_-)^{-1/2}. &(8)\cr} $$
This solution represents a black membrane with event horizon at
$r=r_+$ and inner horizon at $r=r_-$.
Changing coordinates via $y^{-1}=r^{-1}-r_{-}^{-1}$ and taking
the
extremal
limit yields
$$ \eqalignno{e^{2\Phi}&=e^{-2\sigma_3}=\left(1+{y\over
r_-}\right),\cr
ds^2&=-dt^2 + dx_2^2 + dx_3^2 + e^{2\Phi} dy^2,\cr
F&=-{1\over r_-}. &(9)\cr} $$
which is just the supersymmetric domain wall solution$^1$.

Consider the generic toroidal compactification of the heterotic
string, where the four-dimensional theory is given by $N=4$
supergravity coupled to 22 $N=4$ vector multiplets. Then the
Maxwell
field $F_{\mu\nu}$ in $D=4$ (or its dual $\tilde{F}_{\mu\nu}$)
and
the scalar field $\phi$ come from
the $D=10$ $3$-form (or $7$-form) and dilaton plus modulus field of
the
heterotic string (or heterotic fivebrane). Thus, the $D=4$
electric/magnetic
duality can be interpreted as a reduction of $D=10$
string/fivebrane
duality.

Another reduction of string/fivebrane duality is $D=4$
string/string
duality$^1$. The
compactified heterotic string displays a target space duality
$O(6,22,Z)$.
It is also conjectured to display the strong/weak coupling
$SL(2,Z)$ $S$-duality relating the dilaton and the axion, which
is
certainly
there in the field theory limit.
The ``duality of dualities'' suggestion$^{11,1}$ is
that, under string/fivebrane duality, the roles of $S$ and $T$
dualities are
interchanged. The picture that emerges is one in which the
massive
states
of the string correspond to extreme black holes.

We have shown only that these two-parameter
configurations are solutions of the field theory limit of the
heterotic string.
Although the extreme one-parameter solutions are expected to be
exact
to all
orders in $\alpha'$, the same reasoning does not carry over to
the
new
two-parameter solutions. It would be also interesting to see
whether
the generalization of the one-parameter solutions to the
two-parameter solutions can be carried out when we include the
Yang-Mills coupling. This would necessarily involve giving up
the
self-duality condition on the Yang-Mills field strength,
however,
since the self-duality condition is tied to the extreme,
$\sqrt{2}
\kappa {\cal M}_{p+1} =  g_{p+1}$, supersymmetric solutions.
Finally, there is the question of whether these solutions are
peculiar
to the toroidal compactification or whether they survive in more
realistic
orbifold or Calabi-Yau models. Although the actions $S_{1}$,
$S_{2}$
and $S_{3}$ were originally derived in the context of the
torus$^1$,
they
also appear in a large class of $N=1$ supergravity theories.
\vglue 0.6cm
\line{\elevenbf References \hfil}
\vglue 0.4cm
\medskip
\item{1.} M.~J.~Duff and R.~R.~Khuri, {\elevenit Nucl. Phys.}
 {\elevenbf B411} (1994) 473.
\item{2.} M.~J.~Duff, R.~R.~Khuri, R.~Minasian and J.~Rahmfeld,
{\elevenit Nucl. Phys.} {\elevenbf B418} (1994) 195.
\item{3.} G.~W.~Gibbons, {\elevenit Nucl. Phys.}
{\elevenbf B207}
(1982) 337; G.~W.~Gibbons and K.~Maeda, {\elevenit Nucl. Phys.}
{\elevenbf B298} (1988) 741; D.~Garfinkle, G.~T.~Horowitz and
A.~Strominger, {\elevenit Phys Rev.} {\elevenbf D43} (1991)
3140;
R.~Kallosh, A.~Linde, T.~Ortin, A.~Peet and A.~Van Proeyen,
{\elevenit Phys Rev.} {\elevenbf D46} (1992) 5278.
\item{4.} D.~Pollard, {\elevenit J. Phys.} {\elevenbf A16}
(1983)
565; D.~J.~Gross and M.~J.~Perry, {\elevenit Nucl. Phys.}
{\elevenbf B226} (1983) 29;
R.~D.~Sorkin, {\elevenit Phys. Rev. Lett.} {\elevenbf 51}
(1983) 87.
\item{5.} M.~J.~Duff, {\elevenit Class. Quantum Grav.}
{\elevenbf 5}
(1988) 189; A.~Strominger, {\elevenit Nucl. Phys.}
{\elevenbf B343}
(1990) 167; M.~J.~Duff and J.~X.~Lu, {\elevenit Nucl. Phys.}
{\elevenbf
B354} (1991) 129.
\item{6.} E.~Cremmer, S.~Ferrara, L.~Girardello and
 A.~Van Proeyen,
 {\elevenit Nucl. Phys.} {\elevenbf B212} (1983) 413.
\item{7.} R.~R.~Khuri, {\elevenit Phys. Lett.} {\elevenbf B259}
(1991) 261.
\item{8.} R.~R.~Khuri, {\elevenit Phys. Lett.} {\elevenbf B294}
(1992) 325; {\elevenit Nucl. Phys.} {\elevenbf B387}
(1992) 315.
\item{9.} A.~Dabholkar, G.~Gibbons, J.~A. Harvey and F.~R.~Ruiz,
{\elevenit Nucl. Phys.} {\elevenbf B340} (1990) 33.
\item{10.} G.~T.~Horowitz and A.~Strominger, {\elevenit Nucl.
 Phys.}
{\elevenbf B360} (1991) 2930.
\item{11.} J.~H.~Schwarz and A.~Sen, {\elevenit Phys. Lett.}
{\elevenbf B312} (1993) 105; P.~Bin\'etruy,
{\elevenit Phys. Lett.} {\elevenbf B315} (1993) 80; A.~Sen
{\elevenit Int. J. Mod. Phys.} {\elevenbf A9} (1994) 3707.
\vfil\eject
\bye